\documentclass{emulateapj}
\usepackage{graphicx}

\begin{document}

\title{Late time multi wavelength observations of {\em Swift} J1644+5734: A luminous optical/IR bump and quiescent X-ray emission}

\author{A.~J.~Levan\altaffilmark{1}, N.R. Tanvir\altaffilmark{2}, G.C. Brown\altaffilmark{1}, 
B.D. Metzger\altaffilmark{3}, K.L. Page\altaffilmark{2}, S.B. Cenko\altaffilmark{4,5}, P.T. O'Brien\altaffilmark{2}, J.D. Lyman\altaffilmark{1}, K. Wiersema\altaffilmark{2}, E.R. Stanway\altaffilmark{1}, A.S. Fruchter\altaffilmark{6}, D.A. Perley\altaffilmark{7,8}, J.S. Bloom\altaffilmark{9}
}
\altaffiltext{1}{Department of Physics, University of Warwick,
  Coventry, CV4 7AL, UK } 

\altaffiltext{2}{Department of Physics and Astronomy, University of Leicester,
Leicester, LE1 7RH, UK }

\altaffiltext{3}{Columbia Astrophysics Laboratory, New York, NY 10027, USA}

\altaffiltext{4}{Astrophysics Science Division, NASA Goddard Space Flight Center, Mail Code 661, Greenbelt, MD 20771, USA}
\altaffiltext{5}{Joint Space-Science Institute, University of Maryland, College Park, MD 20742, USA}

\altaffiltext{6}{Space Telescope Science Institute, 3700 San Martin Drive, Baltimore, MD 21218, USA} 
\altaffiltext{7}{Department of Astronomy, California Institute of Technology, MC 249-17, 1200 East California Boulevard, Pasadena, CA 91125, USA} 
\altaffiltext{8}{Dark Cosmology Centre, Niels Bohr Institute, DK-2100 Copenhagen, Denmark 0000-0002-4571-2306} 
\altaffiltext{9}{Astronomy Department, University of California, Berkeley, CA 94720-7450, USA} 

\email{A.J.Levan@warwick.ac.uk}


\begin{abstract}
We present late-time multi-wavelength observations of {\em Swift} J1644+57,  suggested to be a relativistic tidal disruption flare (TDF). Our observations extend to $>4$ years from discovery, and show that 1.4 years after outburst the relativistic jet switched-off on a timescale less than tens of days, corresponding to a power-law decay faster than $t^{-70}$.  Beyond this point weak X-rays continue to be detected at an approximately constant luminosity of  $L_X \sim 5 \times 10^{42}$ erg s$^{-1}$, and are marginally inconsistent with a continuing decay of $t^{-5/3}$, similar to that seen prior to the switch-off. Host photometry enables us to infer a black hole mass of $M_{BH}=3 \times 10^6$ M$_{\odot}$, consistent with the
late time X-ray luminosity arising from sub-Eddington accretion onto the black hole in the form of either an unusually optically faint AGN or a slowly varying phase of the transient. Optical/IR observations show a clear bump in the light curve at timescales of 30-50 days, with a peak magnitude (corrected for host galaxy extinction) of $M_R \sim -22-23$. 
The luminosity of the bump is significantly higher than seen in other, non-relativistic TDFs  and does not match any re-brightening seen at X-ray or radio wavelengths.  Its luminosity, light curve shape and spectrum are broadly similar to those seen in superluminous SNe, although subject to large uncertainties in the correction of the significant host extinction.  We discuss these observations in the context of both TDF and massive star origins for {\em Swift} J1644+5734 
and other candidate relativistic tidal flares.

\end{abstract}

\section{Introduction}

Recent years have seen the identification of previously unrecognised populations of extremely long duration $\gamma$-ray transients, visible
for hours to days, compared to seconds or minutes for the well studied populations of GRBs \citep[e.g.][]{levan11,burrows11,gendre13,levan14}. 
These events stretch plausible progenitor models for normal 
GRBs that arise from stellar core collapse, and in particular the longest events have been 
well explained by  the tidal
disruption of stars by supermassive black holes, accompanied by a moderately relativistic outflow, creating a
$\gamma$-ray transient when viewed down the jet \citep{bloom11,zauderer11,burrows11}.

Tidal disruption flares (TDFs) occur when a star strays sufficiently close to a massive black hole that
the tidal force from the hole exceeds the star's self gravity. At this point the star may be
partly or completely disrupted, depending on the pericentre separation and structure of the star itself. 
Roughly half of the disrupted material is expelled, while the remaining bound material is placed
on eccentric orbits, but eventually returns to form an accretion disc around the black hole, powering
a luminous electromagnetic transient, with a black-body spectrum 
expected to peak in the EUV or soft X-ray regime \citep[e.g.][]{rees88}. This process effectively creates a transient
active nucleus, which, unlike most AGN, moves from a quiescent accretion phase through a super-Eddington one, 
and back to quiescence on a timescale of years.

The detection of a TDF provides both a window on accretion physics, and a signal of the presence of a super-massive black hole in an
otherwise inactive galaxy. This is particularly valuable for low-mass galaxies, where direct
confirmation of massive black holes has proved challenging. While some
massive black holes have been identified within dwarf galaxies \citep[e.g.][]{reines11,reines14} 
their interpretation remains uncertain: some lie apparently well off the 
bulge mass -- black hole mass relation \citep[e.g.][]{seth14}, and it is unclear if these
rare examples are representative of other dwarfs (where no activity can be found) or
result from unusual interactions, such as the tidal stripping of more massive galaxies \citep{seth14,reines14b}. 
TDFs can occur, in principle, around all low mass black holes, while they will be
observationally invisible for the most massive systems $M_{BH} > 10^8$ M$_{\odot}$ 
where the tidal radius for a main sequence star lies within the Schwarzschild radius of
the black hole. Thus they provide a particularly powerful probe of the low-mass end
of the nuclear black hole population (e.g.,~\citealt{Metzger&Stone15}), potentially extending down to the
scale of intermediate mass black holes within globular clusters \citep[e.g.][]{ramirez-ruiz09,macleod15a,macleod15b}, and 
offer important constraints on models of galaxy formation and evolution.

However, TDFs themselves are challenging to locate and identify. They are much
rarer than supernovae; they reside in regions of high surface brightness that are
often omitted, or difficult to recover in current transient surveys; and the TDF population itself may exhibit
significant diversity. For example, emission may arise from
the disc, from stream collisions, from an outflow, or 
from an aligned relativistic jet. All of these events may appear observationally distinct, particularly due to
viewing angle effects, and 
hence difficult to distinguish from alternative types of transient. Thus, while
there are many candidate TDFs reported in the literature \citep[see e.g.][for a recent review]{komossa15},
it remains unclear how many really represent tidal flares. Furthermore, the small samples
and various mechanisms
of discovery is such that it is not yet possible to utilise the observed population of candidate
flares to infer the ubiquity and demographics of massive black holes within the nuclei of
different types of galaxy.

A new chapter in this field began in  March 2011 with the discovery of {\em Swift} J1644+57,
a high energy transient unlike any system seem before. 
It originated from the nucleus of a compact galaxy at $z=0.35$
\citep{levan11}, but
its $\gamma$-ray emission persisted for days at the $10^{47}$\,erg\,s$^{-1}$ level (isotropic equivalent luminosity).
It also exhibited an extremely long lived X-ray counterpart \citep{levan11,burrows11},
which remained at a luminosity brighter than  $10^{45}$\,erg\,s$^{-1}$ for more than a year post outburst. 
At first sight these properties do not obviously match the expectations for a TDF. 
Firstly,  the peak luminosity of {\em Swift} J1644+57 exceeds the Eddington limit for
even a $10^{10}$ M$_{\odot}$ black hole. 
It is highly unlikely that this galaxy 
hosts such a black hole since its apparent total stellar mass is less than this value \citep{levan11,yoon15}. 
Indeed, we would not expect to observe  disruptions of main sequence stars around such massive
black holes.
Hence the emission, if isotropic, must be super-Eddington by a factor of 100 or more. 
Secondly, TDFs are expected to be dominated by thermal (or near thermal) emission
with temperature of a few $\times 10^4$~K, while the emission from {\em Swift} J1644+57 was
apparently dominated by a much harder, power-law component enabling its detection by 
the $\gamma-$ray detectors onboard {\em Swift} \citep{bloom11,burrows11}.

Soon after its discovery, it was proposed that these properties could be naturally explained
if {\em Swift} J1644+57 was due to relativistically jetted emission from a tidal disruption event \citep{bloom11a,bloom11}.
In fact, a scenario in which some small fraction of the material from
a TDF was  expelled at relativistic velocities had already been considered,
but primarily from the point of view of possible late time radio emission from
known-TDFs, which may become visible at the point the blast wave is approximately
spherical \citep{giannios11,vanvelzen11}. 
These authors did not consider what may happen when one views directly
down the relativistic jet, since this chance alignment is unlikely. However, this low space density is
compensated (at least to some degree, depending on the beaming angle) by 
the luminosity, providing a much larger horizon over which these events may be seen. 
Given this, \cite{bloom11} suggested that {\em Swift} J1644+57 was in fact 
such an event, effectively a micro-blazar. 
Subsequent precise astrometry \citep{levan11},  the general shape of the X-ray lightcurve, and
the direct measurement of relativistic expansion via radio observations offer
substantial support for this scenario \citep{bloom11,zauderer11}.  
Remarkably, despite seeing none of these events in the first six years of it mission, a second
possible example, {\em Swift} J2058+0516 was uncovered in May 2011 \citep{cenko12}, and
a third, {\em Swift} J1112-8238 \citep{brown15} although only recently recognised, was detected in June 2011. 
It is clear that these events are set apart from previously identified TDFs, maybe because of the impact of viewing
angle \citep{bloom11}, although also perhaps because of unique features of the disrupting system such
as a deeply plunging orbit largely destroying the star \citep{cannizzo11}, or binarity \citep{mandel15}. 

However, alternative hypotheses have also been considered for these systems. 
Specifically, it has been suggested that they could arise from the core-collapse of massive stars, in systems
not unlike those which create long duration gamma-ray bursts \citep{quataert12,woosley12}. 
The basic model to create such events is that material in the outer layers of a rotating massive star has too 
much angular momentum to collapse directly onto the nascent compact object, and instead forms an 
equatorial disc, which feeds the newly formed black hole for a long period of time. 
These events differ from traditional GRBs because it is not the material immediately outside the 
collapsing core forming a relatively short lived disc, but material initially at much larger
distances, creating more massive, long lived accretion events. 
These models were not fully developed until after the initial discovery of {\em Swift}\,J1644+57, and are 
not obviously favoured given the nuclear location of the transient seen in both {\em Swift} J1644+57
\citep{levan11} and {\em Swift} J2058+0516 \citep{pasham15}, and possibly (though not conclusively) 
in {\em Swift} J1112-8238 \citep{brown15}. 
However, to date no conclusive evidence against (or in favour) of them has been found. Interestingly,
similar models have been postulated to explain the origin of the ultra-long GRBs (with durations around $10^4$s \citep{levan14}), where
giant star models have had some success \citep{gendre13,stratta13,nakauchi13,levan14}. Indeed
the recent 
identification of a luminous supernova in the afterglow of the ultra-long GRB~111209A (duration
$\sim 2 \times 10^4$\,s) does apparently demonstrate that core collapse, GRB-like explosions
can occur with durations at least an order of magnitude longer than seen in most GRBs \citep{greiner15}. 

Here we present late time observations of the best studied event, {\em Swift} J1644+57 at
wavelengths from the X-ray to the mid-IR, spanning from 30 days to 4 years after the
detection of the initial outburst. We use these to characterise the light curves and 
host galaxy.  Three striking features are seen, (i) a rapid drop in the X-ray luminosity
500 days post outburst, as also noted by \cite{burrows,levan12,zauderer13}, (ii) an
apparently quiescent underlying X-ray source of luminosity $L_X \sim 5 \times 10^{42}$\,erg\,s$^{-1}$, 
consistent with a low luminosity AGN and (iii) a pronounced bump in the optical/IR light curves, peaking
30--40 days after the initial outburst, with an absolute magnitude of $M_V \sim -22$. We discuss
these properties in light of the expectations of various models for the creation of these extreme
high energy transients.

\section{Observations}
{\em Swift} J1644+57 was discovered by the {\em Swift}-BAT on 28 March 2011 \citep{disc}. Initially classified as a GRB (GRB 110328A) 
the detection of additional bright flares in the following 48-hour period \citep{suzuki11}, and the subsequent discovery of emission in a 4-day window prior to
the initial detection \citep{krimm11} marked it as having exceptionally long $\gamma-$ray emission, persisting for several days \citep[see also][]{levan14}. 
Indeed, a possible detection at $>3 \sigma$ significance was present
in a single day integration $>$1 month before the main trigger \citep{krimm11}. While possibly
a chance noise fluctuation, it is interestingly close to the time of the first trigger that earlier
activity cannot be discounted. 
Although initially suggested to be a Galactic X-ray transient \citep{kennea11}, a redshift of a persistent optical source underlying the
X-ray location revealed a redshift of $z=0.354$ \citep{levan11}, and subsequent monitoring located X-ray, infrared and radio emission
consistent with the nucleus of this galaxy. The early observations have been described in detail \citep{levan11,bloom11,burrows11,zauderer11} and
the source has continued to be monitored by the {\em Swift}-XRT. 
The late time radio afterglow has also received significant monitoring \citep{berger12,zauderer13}. Below
we report the results of ongoing late time optical/IR and X-ray monitoring from both the ground and space. 

\subsection{Further infrared and optical imaging} 
We have continued to monitor {\em Swift} J1644+57 in the IR from the United Kingdom Infrared Telescope (UKIRT) and Gemini-North. A log of our
new photometric observations is shown in Table~\ref{1644_phot}. The UKIRT images were obtained with the Wide Field Camera (WFCAM)
and  reduced through the standard CASU pipeline.  The data were retrieved in
calibrated form from the WFCAM science archive \footnote{http://wsa.roe.ac.uk/}. 
The Gemini-North images were reduced using the standard Gemini IRAF package. Photometric calibration was
performed relative to several 2MASS stars, with the zeropoint tied to the star at
RA=16:44:50.96, DEC =+57:35:31.6 (J=13.121, H=12.798, K=12.727) as in 
Levan et al. (2011), such that the photometric observations should be directly comparable between earlier work and this one. 

We also include in our analysis other published IR photometry from \cite{burrows11}. Observations
taken at similar times provide reasonable agreement with our measured photometry within $\sim 0.1-0.2$ magnitudes, and hence should be on a comparable scale. 
There is no apparent systematic offset that could be applied to reduce this scatter significantly, and so it is likely that the differences in measurements reflect a combination of measurement error (often significant at later times) and true
variation within the source (often significant at earlier times).

\subsection{HST and Spitzer  observations} 

We have also obtained further observations of {\em Swift} J1644+57 with the {\em Hubble Space Telescope} ({\em HST}). 
These observations were obtained in the F606W and F160W bands using the 
WFC3 camera with both UVIS and IR channels, matching the earlier data
presented in \cite{levan11}. 
The images were retrieved from the archive after standard post-processing. The
UVIS observations were corrected for pixel dependent CTE utilising the method of
\cite{anderson}. The images were then drizzled \citep{fruchter2002} onto a common frame, utilising a pixel
scale of 0.025 arcsec/pixel for F606W and 0.07 arcsec/pixel for F160W. The first and last epochs, as well as a subtraction are shown in 
Figure~\ref{mosaic}. To obtain
magnitudes of the counterpart only the final epoch of {\em HST} observations was
subtracted from the earlier data, and the resulting residual measured. The
photometry is shown in Table~\ref{hst_spitzer}, where both transient fluxes
and combined host plus transient magnitudes are listed. To avoid including additional sky noise, which may impair the 
estimation of transient contributions, the combined magnitudes
were measured in an aperture of radius 15 pixels for F160W (1.05\arcsec) and corrected assuming a point-like aperture correction. In practice this
underestimates the true host galaxy magnitude, and so the host galaxy magnitude itself is calculated based on the S\'ersic profile fit to the host galaxy, yielding
a magnitude approximately 0.2 magnitudes brighter. The resulting magnitudes for the host galaxy are comparable to those obtained by \cite{yoon15} from an 
independent analysis of our data. The relatively bright point sources in subtractions were measured in small apertures (2 $\times$ FWHM), and aperture 
corrected, while due to possible galaxy residuals we measured the F606W subtractions in apertures of 0.4\arcsec. We note that as expected 
the choice of aperture size has little impact on our final photometry.

A clear residual is seen in both bands. 
In fact, this is the first detection of transient optical emission in the r-band,
previous detections having only been possible in the z-band and long-wards \citep{levan11}, likely due to the strong 
extinction within the host galaxy. Interestingly, the optical light appears to rise between the first 
two epochs (6.6 and 23 days post outburst) during which time the IR appears to show a decline. 
This is puzzling if both the optical and IR are arising from the same component, and is discussed
further below. 

We can determine the location of the transient within the host galaxy by 
comparing the centroid in the subtracted frames 
with the centre of the host galaxy
in late epoch images, utilising compact sources in the field for astrometric purposes. 
This is best done in the IR since the signal to noise for the transient is much higher, doesn't risk 
any systematic shift due to poorer subtraction of the host galaxy light, and minimises the risk of mis-identifying the centroid due
to differential extinction within the host galaxy. 
We compared our first and last epoch, using 8 sources in common between the two frames for alignment. 
As the first and last images were taken at the same orientation we can utilise a direct shift between the two, rather
than more complex fits (which may underestimate the errors for the small number of sources considered). This
yields an offset of ($0.010 \pm 0.012$)\arcsec, equivalent to a spatial offset of $<60$pc from the centroid of the galaxy. 
Although it has limitations this approach can also be used in the F606W observations, which yields an offset of 
($0.033 \pm 0.010$)\arcsec. This is formally inconsistent with the nucleus at the $3 \sigma$ level, but may be due to 
a combination of the effects described above. 
However, this technique is based on utilising compact sources (predominantly stars) in the field of view, and so proper motion can 
be a significant factor. A new technique, employing cross-correlation with galaxies can improve
this and will be presented separately (Hounsell et al. in prep).

We also observed {\em Swift} J1644+57 with the {\em Spitzer Space Telescope} at four epochs. 
The first three roughly span a year after the outburst, with a final epoch obtained in March 2014 for host subtraction. 
Observations were obtained
in both the 3.6 and 4.5 micron bands. Photometry was performed directly on the PBCD mosaics, and on aligned and 
subtracted images to isolate the transient flux, utilising a 4 pixel (2 native pixel) aperture, and correcting for excluded light. 
The IRAC observations suggest a bright mid-IR outburst, consistent with a highly extinguished source,
which fades by by a factor of 10 over the course of the first year. 
A log of observations and resulting photometry is shown in Table~\ref{hst_spitzer}.

\subsection{Host galaxy spectroscopy}
In addition to the early spectroscopy reported in \cite{levan11} we obtained further optical spectroscopy with Gemini-N/GMOS 
on 23 July 2011 and March 23/April 4 2012. Observations were obtained in the R400 filter, spanning a wavelength range from $\sim 5900-10000$\AA, and
utilising the nod-and-shuffle mode to improve sky subtraction.  The data were reduced via
the Gemini GMOS pipeline appropriate for simple longslit (for our earlier observations) or nod and shuffle (for later data). 
The previously reported emission lines of H$\alpha$, H$\beta$, [OIII] and [OII] \citep{levan11}) remain visible, and no clear evolution is seen. 
In particular, the lines remain narrow
with no sign of the development of broad lines around $H\alpha$, where some recently identified TDF candidate have shown
transient broad features \citep{gezari12,arcavi14}. This is unsurprising given the low level of broad band optical variability in the source,  and 
may be indicative of a lack of broad features, or suggest that the lines seen are from
relatively unobscured star formation within the host galaxy, while any broad line region remains highly obscured.

\subsection{Late time X-ray observations with XMM-Newton and Chandra}

We obtained several epochs of late-time observations of {\em Swift} J1644+57 with both {\em XMM-Newton} and {\em Chandra}. 
A log of these observations with exposure times is shown in Table~\ref{xray}. 
All {\em XMM-Newton} observations utilised the thin filter for both PN and MOS observations. 
{\em Chandra} observations used ACIS-S in very faint mode with the source placed at the default aim point on the S3 chip. 
 
For our {\em Chandra} observations we extracted images from the cleaned event files in the 0.3-10\,keV energy band. 
We then determined count rates in an aperture of 2\arcsec\ radius. 
Although faint, the source is detected in each individual image with between 7--17 source
counts, and within our aperture the background is negligible ($<1$ count expected). Given the small number of counts it is not possible to
determine detailed spectral parameters for our data, although as noted by \cite{berger12} the X-ray photons arise across the energy range, and
are not dominated by soft-photons as would be expected for a thermal blackbody typically thought to underly TDFs. 

The XMM-Newton data were reduced with SAS 14.0.0, using {\sc epchain} and
{\sc emchain} to extract the eventlists. 
All the XMM-Newton observations utilised the thin filter for both pn and
MOS observations; single- and double-pixel events (patterns 0-4) for pn,
and all events up to quadruple pixels (patterns 0-12) for MOS, were
selected. The
eventlists were screened for times of high, flaring background, and an
energy range of 0.3-10 keV was then considered. Source count rates were
extracted using a 10" radius circle centred on the source position, and
corrected for PSF losses caused by the small region size. The background
was estimated from a nearby, larger, source-free region. The numbers given
in Table 3 for the XMM-Newton observations are from the pn datasets in
each case.

We convert the measured X-ray count rates in the 0.3-10\,keV bands into fluxes assuming a simple model 
determined from the fit to the late time X-ray spectra measured by the {\em Swift}-XRT. 
Namely, an absorbed power-law of index $\Gamma=1.99$ and contributions from Galactic and host galaxy absorption 
($N_{H,gal} = 1.75 \times 10^{20}$, $N_{H,host} =  2.07 \times 10^{22}$; \citep{willingale13}). 
We note this does differ in the detail from the fit found by more detailed spectral fitting when the source was brighter, which required an additional 
thermal component providing a few percent of the soft flux. 
However,  the errors associated with the choice of spectrum are small compared to the photon counting errors for the source at this brightness. It is possible
to fit the {\em XMM-Newton} PN observations directly, since the combined observations contains 130 counts (of which approximately half are from the source). 
Doing so, with the absorption fixed to the values determined by the {\em XRT} yields a power-law index of $\Gamma = 1.85_{-0.73}^{0.51}$ (at 90\% confidence),
consistent with the earlier observations and implying no strong hard to soft evolution.

\section{Discussion}
\subsection{Late time X-ray light curve}

The updated X-ray lightcurve of {\em Swift} J1644+57 is shown in Figure~\ref{xlc} on both logarithmic and linear time axes.
Our late-time observations have been supplemented by the ongoing observations with the {\em Swift}-XRT, taken from the
{\em Swift} UK data centre, processed via the techniques described in \cite{evans07,evans09}. As previously noted \citep{levan11,bloom11,burrows11}
the early light curve is dominated by pronounced flaring and variability, which then settles into a steady decay, punctuated by
notable dips, which have been suggested to show some signs of periodicity \citep{saxton12}. The ongoing variability means that
attempts to fit any simple decay model to the data inevitably lead to poor quality fits, although the data from $\sim$100--500 days, if fit with
a single power-law do favour a slope of -5/3 \citep{levan15}. More complex fits could be attempted to investigate the presence or absence
of additional breaks in the light curve, but
these  require some attempt to remove dipping activity, and so are necessarily limited in statistical power. 

The final good detection reported by the {\em Swift}-XRT is at around 500 days, 
with a flux of  $(5.5 \pm 0.8) \times 10^{-13}$\,erg\,s$^{-1}$\,cm$^{-2}$,
based on the stacking of images obtained $\sim 4$ days either side of this midpoint. 
After this, the X-ray flux decreased markedly.
By the time of our {\em XMM}-Newton observations the source had declined by a factor of at least 50 in flux. 
In a factor of $\Delta T / T = 0.08$ in time a fall of a factor 50
corresponds to a decay index of around $t^{-70}$. In practice, the decay was too fast to be resolved since beyond the steep drop-off
XRT observations cannot recover the flux in short exposure times, and there was a significant delay before the
{\em XMM-Newton} and {\em Chandra} observations were scheduled. Hence we conclude that the power-law decay rate was faster
than $t^{-70}$. Assuming we are observing X-ray activity from the base of the jet this suggests that 
activity suddenly shut off, either due to a switch of accretion mode, or the cessation of accretion. 
Given the size of emitting regions at the head of the jet at this late time it is difficult to envision a scenario in
which this shut-off was not due to the cessation of activity close to the base of the jet, since otherwise it would smeared out over a much longer time period. 

It is interesting to note that such rapid cessation of X-ray activity was explicitly predicted in the massive star models of \cite{quataert12}, since
this represents the point at which all of the star has accreted onto the central compact object. Such predictions were not made for jetted-TDF like
events prior to the detection of the rapid drop in {\em Swift} J1644+57, although can potentially be explained via magnetic processes within the disc \citep{tche14}.  In particular, once the black hole accretion rate becomes sub-Eddington and radiatively efficient (geometrically thin), it enters a thermally-dominant accretion state, which are empirically not observed to produce powerful jets in Galactic X-ray binaries (e.g., \citealt{Russell+11}).

After this rapid decay, X-rays of luminosity $L_X \sim 5 \times 10^{42}$ erg s$^{-1}$, continue to be detected until at least April 2015 (day 1500). These X-rays appear to be 
approximately constant in luminosity, with little sign of a decay. 
A fit to the available {\em Swift}-XRT, {\em XMM-Newton} and {\em Chandra} observations
with a constant source is not especially good ($\chi^2$/dof = 13.7/7).
The fit is not improved by allowing for a power-law  model, which gives a
best fit decay $\alpha = 0.5^{+0.7}_{-0.2}$ ($\chi^2$/dof = 10.22/6), with a F-test probability of chance improvement of 20\%. 
However,
these data are dominated by observations immediately after the break, and may contain additional systematic errors from comparison
between three different instruments. If instead we compare the {\em Chandra} count rates then a constant source provides a very
good description ($\chi^2$/dof = 1.27/2), and the power law slope of $\alpha = -0.2^{+0.8}_{-0.4}$ rules out a continuing decay around
$t^{-5/3}$ at $>2.3\sigma$ (and $t^{-4/3}$ at 1.9$\sigma$) .  This is at first sight surprising, since it is reasonable to assume that after the cessation of jet activity we begin
to observe forward shock emission at all wavelengths \citep{zauderer13}. The absence of continued decay of this emission would then suggest 
that these X-rays either don't originate from the forward shock, or that it is somehow continuing to be energised, despite the cessation of jet activity. It is
hence interesting to compare this late time behaviour to the general expectations of differing progenitor models.

In a TDF scenario, once the jet turns off, thermal X-ray emission from the inner accretion disk could be observed (as was originally considered the hallmark signature of TDFs; e.g., \citealt{rees88}).  For stellar tidal disruption by a black hole of mass $\sim 10^{6}M_{\odot}$, the fall-back time of the most tightly bound tidal debris is $t_{\rm fb} \sim$ 1 month, similar to the duration of peak hard X-ray activity in {\em Swift} J1644+57 and J2058+05.  If the black hole accretion rate is assumed to faithfully track the mass fall-back rate $\dot{M}$, then the thermal accretion luminosity at some time $t$ later is approximately given by
\begin{eqnarray}
L_{\rm X} \approx \eta \dot{M}c^{2} \approx \eta\frac{M_{\star} c^{2}}{3t_{\rm fb}}\left(\frac{t}{t_{\rm fb}}\right)^{-5/3} \nonumber \\
\approx 3\times 10^{43}\,{\rm erg\,s^{-1}}\left(\frac{\eta}{0.1}\right)\left(\frac{M_{\star}}{0.5M_{\odot}}\right)  \nonumber \\  
\times \left(\frac{t_{\rm fb}}{\rm month}\right)^{2/3} \left(\frac{t}{\rm 1000\,d}\right)^{-5/3}, \nonumber \\
\label{eq:Lx}
\end{eqnarray}
where $\eta$ is the accretion efficiency and $M_{\star}$ is the mass of the disrupted star.  To order of magnitude, the predicted luminosity at 500-1000 days is similar to that observed in J1644+57 after the steep drop (once a bolometric correction is included).  However, the predicted $\propto t^{-5/3}$ decay is steeper than the observed light curve between 500 and 1000 days.   A dimmer and flatter light curve than predicted by equation~\ref{eq:Lx} could be explained if the black hole accretion rate after the jet shut-off no longer tracks the mass fall-back rate, due to the viscous spreading of the disk (\citealt{Cannizzo+90,Shen&Matzner14}).  Such a transition from rapid to slow processing by the disk is naturally instigated by the sudden and large increase in the viscous timescale $\propto H^{-2}$, once the disk scale-height $H$ shrinks following the sub-Eddington state transition (\citealt{Shen&Matzner14}). However, the apparently relatively hard X-ray spectrum after the rapid decay is not in keeping with the very soft thermal spectrum expected in TDFs, and so it seems less likely that this is the observed origin of the late time X-ray emission.

In the case that all the material from a collapsing star has been accreted \citep{quataert12} it seems unlikely that an essentially quiescent source would persist. 
One possibility is that some level of ongoing accretion may occur from the
dense region in which the SN occurred, although the luminosity is orders of magnitude larger than possible from either Bondi-Hoyle accretion 
in a giant molecular cloud, or from accretion from a companion star. 
Indeed, the luminosity of $\sim 5\times 10^{42}$ erg s$^{-1}$ remains
$\sim$3 orders of magnitude larger than possible from a stellar mass black hole, and would require both a continued high accretion rate, and a significant
degree of beaming unless the supernova had been from an extremely massive star that had created an exceptionally massive black hole \citep[e.g.][]{zwart04}. 

Finally, it is possible that the late time X-rays represent ongoing AGN activity, separate to the transient outburst. 
The X-ray luminosity itself would be fairly typical for a low-luminosity AGN, however, the host galaxy would be unusual in this case since
the majority of AGN are hosted in rather more luminous galaxies. 
This is illustrated in Figure~\ref{lxmopt} which, following \cite{levan11}, shows the comparative luminosity evolution of {\em Swift} J1644+57 
in the X-ray luminosity against the optical/IR absolute magnitude plane. 
The track of the counterpart of {\em Swift} J1644+57 is shown at several 
characteristic times, and shows that it evolves from extreme X-ray luminosity through to rather fainter luminosities in both the optical/IR and X-ray. 
However, at late times it does not fall within the locus of X-ray emitting galaxies, either of local galaxies harbouring relatively quiescent 
black holes, or of more luminous AGN.  For example, in the comparison of   \cite{pineau11} of SDSS with 2XMM, only a handful of matches 
have optical absolute magnitudes fainter than -19, and in most of these galaxies the X-ray luminosity 
is sufficiently low ($10^{38-40}$ erg s$^{-1}$) that discrete X-ray emission from binaries etc. could be responsible for the observed flux. 
Indeed, the optical absolute magnitude of the host galaxy of 
{\em Swift} J1644+57 of $M_V \sim -18.5$ is fainter than the cases of Heinze 2-10 \citep{reines11} or Mrk 709 \citep{reines14}, both
nearby dwarf galaxies thought to harbour massive black holes. 
Thus, despite the apparent plateau in X-ray luminosity, this argues against the presence of a standard AGN within
the host galaxy, as supported by the absence of obvious AGN features in either optical spectroscopy (see above) or late time radio follow-up \citep{zauderer13}. 
Further
X-ray observations over increasingly long time periods should ultimately offer a sensitive test of any variability within the source. 

\subsection{Optical/IR lightcurve}
A striking feature of the optical/IR light curves is the presence of an apparent upturn to a peak in the light curve around 30 days after the outburst. 
Initially the plateauing seen at these times was assumed to be due to the source fading into its host galaxy light, 
but later observations clearly demonstrate further fading by a factor of $>3$ from this time.
There is significant point to point scatter in the IR observations at many epochs, possibly due to intrinsic variation in the source on short timescales. 
Direct comparison of observations taken with the same instrument and telescope combination implies that this variability is real, at least at early times. 
There is also likely to be some scatter due to slight systematic differences in the photometry between our own and those
reported by  \cite{burrows11}.  
This means that as with the X-ray, simple fits to the data do not yield high quality fits, and will provide only an
approximation of the true behaviour. 
However, the host subtracted K-band data can be described by a multiply broken power-law as shown in Figure~\ref{optlc}. 
The counterpart declines with $\alpha_1 \approx 1.3$ (where $F_{\nu} \propto t^{-\alpha}$), the rises with $\alpha_2 \approx -0.7$ to a peak 
at 30 days. From this point a decline with $\alpha_3 \approx 0.8$ describes the final fading into the host galaxy, although there are
significant errors on the late time points due to the uncertainty in host subtraction.  
This crude model of three power-law segments also provides a reasonable
fit to the H- and J-band observations if an arbitrary offset is applied (see Figure~{\ref{optlc}). If this offset is scaled to provide a good match to the early data ($<$10 days post
burst) then it significantly under predicts the strength of the bump in the H and J bands. This suggests that the bump does not have the
same underlying spectral energy distribution as the earlier counterpart, and is much bluer with relatively weak IR emission. 

The {\em HST} observations provide the best measurements of this bump since they can cleanly be subtracted for
host contribution without the need for PSF matching, or differences between cameras or filters. However, the 
{\em HST} observations also provide extremely poor temporal sampling. Nonetheless, it is striking to note that
the F606W optical observations show an apparent rise between 6 and 23, with the 23 day flux $\sim 1.5$ times brighter than
at day 6, while the
IR light at 23 days is 0.9 times as bright as at 6 days (see Figure~\ref{sncomp}). This offers further evidence that the bump is a separate
feature, rather than a simple, achromatic rebrightening. The {\em HST} observations also suggest that
at later times the decay cannot be well fit as a single power-law decay, although this is again based on 
very small sampling (3 points per band). 

In Figure~\ref{optsed} we show the evolution of the spectral energy distribution of {\em Swift} J1644+57. It can be seen to be extremely red, as previously noted. 
Its SED, combined with the significant X-ray column density favours an optical extinction in the region $1.5 < E(B-V) < 2$ \citep{levan11,bloom11,burrows11}. 
To highlight the possible impact
of extinction we then also plot the SED corrected for a maximal extinction  of $E(B-V)=2$, assuming a Milky Way extinction law, 
although since none of our wavelengths are close to the rest-frame 2175\AA~ bump the choice of extinction law has minimal impact on the
correction. The peak of the bump at 30 days has an absolute magnitude of
$M_B \sim -22$ for a maximal extinction, comparable to the peak magnitude in the K-band (which is far less affected by host extinction).

\subsection{The origin of the optical bump} 

Perhaps the most plausible explanation is that the optical bump originates as the hot thermal component of the tidal flare. 
Such components are typically those expected based on non-relativistic models \citep[e.g.][]{rees88}. 
This peak luminosity occurs well after the disruption itself, since the peak accretion rate is after the return of the most bound debris. 
Indeed, numerical models of mass return suggest that luminous UV flares may peak on timescales of $\sim 20-50$ days at optical and UV
wavelengths \citep{lodato11} with luminosities rather similar to those of normal SNe. 
This is broadly borne out by observations of candidate disruptions to
date, with many of the most promising candidates showing such rises. 
However, suggested examples of TDFs actually show a surprisingly large
variation in their properties. Some peak early and very bright ($M_V < -20$ and rise times of a few days, e.g. PTF10iya \citep{cenko11}) while those
with much longer life spans are also significantly fainter ($M_V > -18$ with rise times of 20-50 days). 
There are no examples which apparently match the
energy output for {\em Swift} J1644+57, although there remains significant uncertainty about both the extinction and the contribution of any non-thermal component. 
We plot the light curves of {\em Swift} J1644+57 (after subtraction of the host contribution and correction for host extinction) against those of candidate TDFs in Figure~\ref{sncomp}. 
Unfortunately, such a comparison is non-trivial since {\em Swift} J1644+57 is predominantly observed in the rest-frame IR, while 
the thermal flares are strong UV and optical emitters. 
Therefore, the poorly sampled optical light curve of {\em Swift} J1644+57 (cyan line in Figure~\ref{sncomp}) is probably the best comparison with known
examples. 
Such a comparison is also complicated since the origin of many suggested TDFs remains uncertain, for example some may be unusual SNe, 
others due to partial disruption \citep[e.g.][]{chornock14,holoien14}, or the disruption of unusual stars \citep[e.g.][]{gezari12}. 

Another possibility is that the re-brightening is due to the optical/IR contribution of the second synchrotron component identified by \cite{berger12}. This
peaks at radio wavelengths at $\sim 100$ days, although plausible synchrotron models could result in an earlier peak for the optical/IR emission (as seen 
in GRBs for example \citep{sari98}), depending on the location of spectral breaks. This would have the appeal of representing the manifestation of a feature
known at other wavelengths, and might also explain the relatively high polarisation ($7.4 \pm 3.5 \%$) seen in the IR, 17 days after the outburst, as the bump is beginning to dominate
\citep{wiersema12}. Indeed, as noted by \cite{decolle12} the delay between viewing energy injection at the base of the jet in X-ray's and radio emission from the jet-head is naturally expected in models of jetted TDFs, and this ``lag" in which the optical/IR peak is between these two extremes may have some appeal.  
However, the parameters would necessarily require some tuning to provide the brightening without the presence of any moving spectral breaks in the optical/IR, since while the relative strength of the bump emission varies with wavelength, the shape of the bump is broadly similar.  The bump colours would also be unusually blue -- corrected for host extinction the spectrum would follow 
$F_{\nu} \sim \nu^{2}$ or steeper, much bluer than expected for GRB-blast waves in this wavelength regime. The polarisation measurement could also represent
underlying asymmetry in the source, as is seen in some SNe \citep[e.g.][]{patat11}, while its intrinsic value is significantly uncertain since interstellar polarisation within
the host could also play an important role \citep{wiersema12}. 
To date there do not exist polarimetric observations of the thermal components of TDF flares, and so this cannot be compared directly.

An alternative hypothesis is that the optical bump could be due to reverberation of the X-ray light. \cite{yoon15} claim that the morphology of the 
optical is similar to that of the X-ray, but with a delay of $\sim 15$ days. While this does not appear the case in a detailed comparison (for example the
X-ray rise is  rapid while the optical/IR rise apparently takes place over the timescale of several days) it is possible that a prompt
injection of energy in the X-ray could be smoothed out should the reverberating material be spread out at an average distance of $\sim 15$ light
days from the central engine. While the lags to the broad line region can be of this size  \citep{peterson04} simultaneous optical/X-ray monitoring of AGN
typically yields much smaller lags ($\sim 1$ day) between X-ray and optical emission \citep[e.g.][]{breedt09,breedt10}, while lags due to processes within 
the disc are also short ($\sim 1$ day), and should increase with increasing wavelength \citep{mchardy14}. Hence the properties
of the light curves do not naturally match the expectations of reverberation seen in AGN, and would require an unusual, pre-existing 
AGN-like geometry to exist within the host. On the other hand, the unexpectedly high optical luminosities and low
effective temperatures of many optically-selected TDFs have also been
attributed to "reprocessing" of the inner disk emission by debris from the
merger, either bound debris still returning to the BH \citep{Guillochon13}
or an unbound outflow from the accretion disk (e.g., Strubbe \&
Quataert 2011; \citealt{Metzger&Stone15}).

The other class of astrophysical transient that can reach such extreme luminosities are the superluminous supernovae (SLSNe) \citep[e.g.][]{gal-yam12}.
 These events peak at magnitudes of $M_V < -21$, and have slow rise times of 30-100 days, followed by slow decays. 
The peak luminosity of {\em Swift} J1644+57 
is comparable to these events, and given the uncertainty in both the explosion date of SLSNe, and the true ``trigger" time for {\em Swift} J1644+57 it
is possible to obtain a reasonable match in both light curve shape and luminosity. 
For the case of $E(B-V) =2$ the luminosity would be amongst the 
highest for SLSNe, although the recent discovery of the most luminous SLSNe ASASSN-15lh would be comparable (in fact, it should also be noted that
ASASSN-15h is apparently coincident with the nucleus of its host galaxy \citep{dong15}, as is the second brightest SLSNe, CSS 100217, \citep{drake11}, perhaps offering
further hints of similarities between classes of astrophysical transient	). 
Given the uncertainties in host extinction one can also
find a reasonable match in terms of spectral shape between hydrogen poor SLSNe, and {\em Swift} J1644+57 (see Figure~\ref{optsed}). 

At first sight, the strong simultaneous X-ray
emission would appear to rule out an SLSNe origin, however two recent developments may be important in this regard. 
Firstly, the apparently normal hydrogen poor SLSNe, SCP 06F6 has a strong X-ray detection $>100$ days after its discovery, 
with a luminosity very similar to that of  {\em Swift} J1644+57 at the same epoch \citep{gaensicke09,levan13}. 
X-ray observations were not obtained of SCP06F6 until very late, but it is possible that it is due to
jet-like emission that could have been persistent but undetected over a long period in a system similar to {\em Swift} J1644+57. 
Although it is also possible that the X-ray detection of SCP 06F6 was due to a shorter breakout 
of magnetar emission \citep{levan13}, and the possibilities cannot be distinguished between with the paucity of earlier X-ray observations.
Secondly, in the case of one ultra-long GRB, GRB 111209A \citep{levan14,gendre13} there has recently been the identification of a luminous supernova
signature \citep{greiner15}, indicating that one can simultaneously observe strong X-ray emission and a luminous SNe bump. If these
SNe are in fact powered by either a black hole or long-lived magnetar central engines then one might expect to sometimes observe them
down a jet-axis, in which case events like {\em Swift} J1644+57 or GRB 111209A could be observed. Motivated in part by these results \cite{metzger15} have
shown that the full variety of luminous SNe and extremely long-lived high energy transients can be explained (although not necessarily uniquely) 
by magnetars with differing magnetic fields and spin
down times, extending the suggestion by \cite{mazzali14} that most GRBs can be explained by such a mechanism. 
Indeed, they note that this model would naturally predict the luminosity of {\em Swift} J2058+0516. The case of {\em Swift} J1644+57 would then
also fit on the extrapolation of these models.

Indeed, it is instructive to consider {\em Swift} J2058+0516 in this regard, since it exhibited similar 
high energy properties to {\em Swift} J1644+57, but lacked the heavy extinction. 
Thus we might expect to be able to test any SN hypothesis, especially as the redshift was almost identical to the SLSNe SCP 06F6. 
In this case the luminosity of the optical afterglow was comparable to SLSNe, and the inferred temperature ($T \sim 2 \times 10^4 K$, \cite{pasham15}) was similar 
to both GRB 111209A/SN2011kl and ASASSN-15lh \citep{greiner15,dong15}. However, there was only rather minimal 
evidence of any optical rise (although observations started late) and optical spectroscopy did not yield any sign 
of the strong absorption features seen in most SLSNe. 
This casts some doubt on any model linking events such as 
{\em Swift} J1644+57 and {\em Swift} J2058+0516 with stellar core collapse, although it should equally 
be noted that in the case of GRB 111209A/SN2011kl \citep{greiner15} the high ejecta velocities diluted 
any absorption features such that they were not obvious in the observed spectra. The final case of {\em Swift} J1112-8238 \citep{brown15} unfortunately does not
yield such strong constraints due to rather patchy follow-up, although the absolute magnitude of the transient of $M_B \sim -21.4$, 20 days after the BAT detection is
in keeping with the absolute magnitudes seen in both {\em Swift} J1644+57 and {\em Swift} J2058+0516.

Finally, the observed rates of different events could potentially provide some discrimination between progenitor models. Before correction for beaming,
GRBs likely show a volumetric rate of a few Gpc$^{-3}$, corrected for likely beaming this becomes $\sim 300$ Gpc$^{-3}$ \citep[e.g.][]{Kanaan}, 
rather comparable to the rate of SLSNe \citep{quimby13}. The rate of {\em Swift} J1644+57-like transients, or ultra-long GRBs is significantly lower than the GRB rate, although poorly constrained
given the small population observed, and observational biases against their detection as long lived, low peak-flux events \citep{levan14}. 
\cite{brown15} estimate a rate of  $3 \times 10^{-10}$ galaxy$^{-1}$ yr$^{-1}$ for {\em Swift} J1644+57-like events. Accounting for
biases in their detection could give an order of magnitude larger rate, with a similar boost given if shorter events, such as the ultra-long GRBs are included \citep{levan14}. 
Given the volume density of
galaxies in the relatively local Universe (or more specifically massive black holes) is $\sim 10^{-2} - 10^{-3}$ Mpc$^{-3}$ the inferred volumetric rate of the {\em Swift} J1644+57 like events is
$\sim 3 \times 10^{-3}$ Gpc$^{-3}$ yr$^{-1}$, or allowing for the various selections against their discovery perhaps as high as $\sim 0.1$ Gpc$^{-3}$ yr$^{-1}$. Hence, even with very small beaming angles (e.g. the factor of $\sim 10^{2-3}$ needed to bring the observed luminosity below the Eddington limit
for a $10^{7-8}$ M$_{\odot}$ black hole) such jets need only be launched from a small fraction of SLSNe. This would explain why evidence for
their existence in X-ray monitoring of SLSNe is rare to date \citep{levan13}. Equally, these rates are significantly below the rates of tidal disruption flares, whose canonical rate
of $1 \times 10^{-5}$ galaxy$^{-1}$ yr$^{-1}$ is 5 orders of magnitude higher than that of the relativistic counterparts. As noted by \cite{cenko12} and \cite{brown15} it is therefore
unlikely that any significant fraction of TDFs could launch such powerful relativistic jets as seen in {\em Swift} J1644+57 and other examples.  Overall, the rate arguments suggest that these very-long duration transients could arise from some small subset of either TDFs or SLSNe.

\subsection{Host galaxy properties}
After the X-ray break it is likely that the observed flux in all bands is now dominated by the host galaxy,
affording us the opportunity to investigate it in more detail than previously possible. Indeed, this is supported by
the analysis of \cite{yoon15} who attempt to fit a point source onto the host, concluding that at later times the point source
contribution is minimal. 
The galaxy is detected in 12 photometric bands from 0.45--4.5 microns (see Table~\ref{host_tab}), with limits at both shorter and longer 
wavelengths (although in practice these limits are not yet constraining). 
From this we can derive the physical properties of the host galaxy based on template fitting
to the available spectral energy distribution (SED) shown in Figure~\ref{galsed}. 
Considering the Binary Population and Spectral Synthesis (BPASS) library of models \citep{eldridge09}, 
we find the SED to be well reproduced by a relatively old dominant stellar
population (age $= 3.2 \times 10^9$ years), although the emission lines clearly indicate the presence of a younger population as well (Fig.~\ref{galsed}). 
Importantly, the fitting also provides a much more robust determination of the stellar mass than was previously possible, 
since earlier attempts were significantly contaminated at red wavelengths by transient light. 
Specifically we find a stellar mass of $M_* = 5.5 \times 10^9$ M$_{\odot}$. 
This value is somewhat larger than that found by \cite{yoon15}
from their more detailed study ($M_* = 1.38^{+0.48}_{-0.27} \times 10^9$). 
However, this may be explained by the use of differing spectral models, and our
use of later time {\em Spitzer} observations, free from transient contamination. 
This stellar mass can be used to infer an approximate
mass for the central black hole. 
\cite{Scott_etal} 
find that for core-S\'ersic profiles the scaling is roughly linear ($M_{BH} \propto M_*^{0.97 \pm 0.14}$),
but for galaxies with low masses (they define low to be $M_* < 3 \times 10^{10}$ M$_{\odot}$) they find a much steeper relation
of $M_{BH} \propto M_*^{2.22 \pm 0.58}$. Under the assumption that the galaxy stellar mass is equal to its spheroid mass (which seems
a reasonable assumption given the surface brightness profile, see below) 
the implied black hole mass is then $M_{BH} \sim 3 \times 10^6$ M$_{\odot}$, which could be 
taken as an upper limit on the likely BH mass. 

Despite its luminosity appearing very similar to the LMC ($M_B \sim -18$), the morphological properties of the host of {\em Swift} J1644+57 are 
rather different. The core of the galaxy is barely resolved by the HST IR observations, although is reasonably resolved
in the optical. The galaxy has little ellipticity $e\approx0.1 $ and is very concentrated, with 
$R_{20,50,80} =0.077,0.184,0.388 \arcsec = 0.39,0.92,1.95$\,kpc at $z=0.354$. 
Its surface brightness profile is well fit with a S\'ersic fit
with $n=4$ (i.e. a de Vaucouleurs profile) in both the optical and IR, 
suggesting it is dominated by a spheroidal component (see also \cite{yoon15}. However
a subtraction of a rotated image does reveal some asymmetry with a knot-like structure extending $\sim$ 0.1\arcsec~ from the galaxy nucleus, but
interestingly including the location of the transient. These
are potentially the star forming regions creating emission lines, and lead to a formal concentrated asymmetry measure of 
$C \approx 3.5$, $A \approx 0.1$, placing the host in a region of in the concentration asymmetry plane similar to many GRB hosts
\citep{conselice05}. 

\section{Summary}

We have presented multi-wavelength observations of {\em Swift} J1644+57, continuing for $>$3 years after its initial detection. At this stage the
observed light at X-ray, optical and infrared wavelengths appears to be dominated by quiescent emission. In the case of the optical/IR this
is likely the host galaxy. In X-rays, an apparently persistent source of luminosity $L = 4 \times 10^{42}$ erg s$^{-1}$ either represents a slowly
declining phase of the counterpart, or an underlying low luminosity AGN. The presence of an AGN in a tidal disruption event is not unprecedented, in
particular the recent ASASSN-14li appears to arise from a pre-existing radio galaxy, and indeed the coincidence of the source with an apparently
active nucleus may increase the confidence in which it can be assigned to a tidal disruption flare. 

More puzzling is the nature of the optical and IR emission. A strong, luminous bump at $\sim 30$ days with an absolute magnitude of $M_B \sim -22$
is not well matched by the thermal bumps seen in other tidal flares, since it is much more luminous. The bump is more pronounced than seen in the case of
the other candidate flares {\em Swift} J2058+0516 and {\em Swift} J1112-8238, although this may be due to the earlier initiation of observations in the case
of {\em Swift} J1644+57. For {\em Swift} J1644+57 observations were taken within hours of the BAT trigger, and within at most a few days of the clear onset of activity, 
in the case of {\em Swift} J2058+0516 and {\em Swift} J1112-8238 the first optical observations took place $>$10 and 20 days after the BAT detections respectively, meaning
that any rise could have been missed. In all three cases the luminosity of the counterpart is brighter than $M_B \sim -21$.

The properties of these bumps may represent extreme versions of the thermal flares from TDFs. In the case of {\em Swift} J2058+0516 the inferred temperature is
comparable to those found for thermal TDF flares, and the soft X-ray components \citep{levan11,burrows11} may also be consistent with those expectations, although
the inferred temperatures of the X-ray black bodies are much higher than inferred from the optical. Given the
apparent differences in total energetics in relativistic TDFs and thermal events it might be less surprising that the thermal bumps are also different, and may reflect
differences in the stars being accreted (differences in mass, radius, magnetic field, binarity etc \citep[e.g.][]{krolik11,mandel15}. 
Alternatively, it may be that these 
events are not in fact from tidal disruption flares but from luminous supernovae explosions. In this case they may arise when a luminous SN launches a relativistic 
jet on collapse, in which case they would be GRB-like events arising from some subset of SLSNe, as normal long duration GRBs arise from some small subset of 
SN Ic. The observed rate of SLSNe are comparable to those of GRBs \citep{quimby13}, while the rates of the very long transients, even allowing for beaming factors
of 100-1000, are much lower, implying that visible high energy transients associated with SLSNe would be rare, even compared to the SLSNe rate.  \citet{Metzger&Stone15} develop a model for the optical TDF emission which is qualitatively similar to those developed for engine-powered SLSNe (i.e., reprocessing of central engine energy by approximately a solar mass of outflowing matter; e.g., \citealt{Dexter&Kasen13}), highlighting the challenges of distinguishing TDFs and core collapse events based on their optical light curves alone.

Further diagnostics are clearly needed to form firm conclusions. There are likely to be three routes through which this can come. The first is via spectroscopy of the
bumps in any further examples. High quality spectroscopy, allied to detailed modelling can yield diagnostics even in the case of relatively weak or featureless
spectra, as recently demonstrated in the case of the ultra-long, and luminous supernovae pairing GRB 111209A/SN2011kl \citep{greiner15}. The unique identification
of features expected in luminous SNe (e.g. turn-off due to line blanketing, absorption lines seen in SLSNe) or TDFs (e.g. blue shifted narrow lines from 
streams \citep{strubbe11}) would then provide a clinching argument as to the
origin of the bumps in the longest high-energy transients. A second route arises through studying the locations of the transients within their hosts. {\em Swift} J1644+57
clearly arises very close to the galactic nucleus, and {\em Swift} J2058+0516 is also consistent with the nucleus of a much fainter galaxy \citep{pasham15}. In
the case of GRBs approximately 1/6 of examples are consistent with a galaxy nucleus \citep{fruchter06,svensson10}, this number may be lower for SLSNe \citep{lunnan15}
although the origin of SLSNe in the nuclei of galaxies may be ambiguous \citep[e.g.][]{dong15}. Further examples, all in the nucleus of their hosts would rapidly remove
any SNe model from consideration. Finally, we can also consider the host galaxy more globally. TDFs can be observed in quiescent, non-star forming galaxies while 
SLSNe are thought to arise from massive star collapse \citep{gal-yam12}. Although magnetars similar to those suggested to power SLSNe can be formed via accretion
induced collapse \citep{usov92,levan06}, the lack of any significant remnant to re-energize via spin-down means that the presence of an extremely long event within an
quiescent elliptical galaxy would rule out SNe models, and strongly favour an origin as a relativistic tidal flare. Since a reasonable fraction of candidate tidal disruptions arise from
passive systems \citep[e.g.][]{arcavi14} such a test should be possible with only a handful of additional examples.

\section*{Acknowledgements} 
AJL, NRT, KW, PTO thank STFC for support. KLP thanks the UK Space Agency. 
We thank Matt Mountain, Harvey Tannenbaum and Norbert Schartel and the teams from STScI, CXC and ESAC 
for the approval and rapid scheduling of DDT observations with {\em HST}, {\em Chandra} and {\em XMM-Newton} respectively.  

Based on observations made with the NASA/ESA {\em Hubble Space Telescope}, obtained [from the Data Archive] at the Space Telescope Science Institute, which is operated by the Association of Universities for Research in Astronomy, Inc., under NASA contract NAS 5-26555. These observations are associated with {\em HST} programs GO 12447, 12378 and 12764. 

The scientific results reported in this article are based to a significant degree on observations made by the Chandra X-ray Observatory. The observations
reported are from programme numbers 12900486, 13708437 and 15700509.

Based on observations obtained with XMM-Newton, an ESA science mission 
with instruments and contributions directly funded by 
ESA Member States and NASA.

Based on observations obtained at the Gemini Observatory, which is operated by the 
Association of Universities for Research in Astronomy, Inc., under a cooperative agreement 
with the NSF on behalf of the Gemini partnership: the National Science Foundation 
(United States), the National Research Council (Canada), CONICYT (Chile), the Australian 
Research Council (Australia), Minist\'{e}rio da Ci\^{e}ncia, Tecnologia e Inova\c{c}\~{a}o 
(Brazil) and Ministerio de Ciencia, Tecnolog\'{i}a e Innovaci\'{o}n Productiva (Argentina).

This work made use of data supplied by the UK Swift Science Data Centre at the University of Leicester, funded by the UK Space Agency.

\begin{deluxetable}{llllllll} 
\tablecolumns{8} 
\tablewidth{0pc}
\tablecaption{Ground-based photometric observations of {\em Swift} J1644+57} 
\tablehead{ 
\colhead{MJD-obs} & \colhead{$\Delta T$} & \colhead{Telescope} &\colhead{Band} & \colhead{Mag}  }
\startdata
55775.26 & 126.73 & UKIRT/WFCAM & K & 20.72 $\pm$ 0.07   \\
55775.26 & 126.76 & UKIRT/WFCAM & J & 21.52 $\pm$ 0.03  \\
55841.24 & 192.70 & UKIRT/WFCAM & K & 20.53 $\pm$ 0.17  \\
55843.20 & 194.66 & UKIRT/WFCAM & K & 20.94 $\pm$ 0.08  \\
55843.24 & 194.70 & UKIRT/WFCAM & J & 21.84 $\pm$ 0.23  \\
55844.20 & 195.66 & UKIRT/WFCAM & H & 21.57 $\pm$ 0.20\\
55999.64 & 351.10 & Gemini/GMOS & r & 22.49 $\pm$ 0.02  \\
55999.66 & 351.12 & Gemini/GMOS & z & 21.94 $\pm$ 0.02 \\
56049.45 & 400.91 & Gemini/NIRI & K & 21.42 $\pm$ 0.04  \\
56049.47 & 400.93 & Gemini/NIRI & H & 21.83 $\pm$ 0.09  \\
56049.49 & 400.95 & Gemini/NIRI & J & 21.99 $\pm$ 0.11 \\
56108.39 & 459.85 & Gemini/NIRI & J & 21.90 $\pm$ 0.06   \\
56108.43 & 459.89 & Gemini/NIRI & H & 21.63 $\pm$ 0.04 \\
56108.46 & 459.92 & Gemini/NIRI & K & 21.30 $\pm$ 0.05 \\
56211.20 & 592.66 & UKIRT/WFCAM & K & 21.22 $\pm$ 0.12 \\
\enddata
\tablecomments{Magnitudes are not host subtracted.
This table supplements the photometry given in \cite{levan11}.}
\label{1644_phot}
\end{deluxetable}

\begin{deluxetable}{llllllll} 
\tablecolumns{8} 
\tablewidth{0pc}
\tablecaption{Late-time space-based optical/IR/mIR observations of {\em Swift} J1644+57} 
\tablehead{ 
\colhead{Date-obs} & \colhead{MJD-obs} & \colhead{$\Delta T$} & \colhead{Telescope} &\colhead{Band} & \colhead{Exptime} & \colhead{Transient flux}  & \colhead{Mag} \\
& & \colhead{(days)} & & & \colhead{(s)} &  \colhead{ ($\mu$Jy)} & \colhead{(Host+OT)}}
\startdata
2011-04-04 & 55655.13 & 6.59 & HST & F160W & 997 & 10.47 $\pm$ 0.04 & 20.76 $\pm$ 0.01 \\
2011-04-04 & 55655.14 & 6.60 & HST & F606W & 1260  & 0.129 $\pm$ 0.023 & 22.82 $\pm$ 0.03  \\
2011-04-20 & 55671.56  & 23.02 & HST & F160W &  997  & 9.545 $\pm$ 0.04 & 20.83 $\pm$ 0.01\\
2011-04-20 & 55671.57 &  23.03  & HST & F606W & 1260 & 0.185 $\pm$ 0.021 & 22.76 $\pm$ 0.04 \\
2011-08-04 & 55777.26 &  128.72 & HST & F160W & 1412 & 3.39 $\pm$ 0.04 & 21.29 $\pm$ 0.01\\
2011-08-04 & 55777.27 &   128.73 & HST & F606W & 4160 & 0.09 $\pm$ 0.015  & 22.89 $\pm$ 0.02\\
2011-12-02 & 55897.70  & 249.16 & HST & F160W &  1209 & 3.13 $\pm$ 0.03 & 21.35 $\pm$ 0.01\\
2011-12-02 & 55897.68 &  249.14  & HST & F606W & 1113  & 0.004 $\pm$ 0.020 & 22.92 $\pm$ 0.05\\
2013-04-12 & 56394.30  & 745.76 & HST & F160W & 2812 & - & 21.73 $\pm$ 0.01 \\
2013-04-12 & 56394.44 &  745.90  & HST & F606W & 2600  & - & 22.93 $\pm$ 0.03 \\

2011-04-28 & 55679.98 & 31.44 & Spitzer & 3.6 & 480 & 58.00 $\pm$ 1.76 & 19.39 $\pm$ 0.02 \\ 
2011-04-28 & 55679.98 & 31.44 & Spitzer & 4.5 & 480  & 72.96 $\pm$ 1.75 & 19.18 $\pm$ 0.02  \\
2011-10-31 & 55865.02 & 216.48 & Spitzer & 3.6 & 480 &  4.02 $\pm$ 1.86 & 21.30 $\pm$ 0.12 \\ 
2011-10-31 & 55865.02 & 216.48 & Spitzer & 4.5  & 480 & 6.95 $\pm$ 1.55 & 21.10 $\pm$ 0.10\\
2012-02-24 & 55981.54 & 333.00 & Spitzer & 3.6  & 480  & 3.31 $\pm$ 1.68 & 21.37 $\pm$ 0.09 \\
2012-02-24 & 55981.54 & 333.00 & Spitzer & 4.5  & 480  & 3.63 $\pm$ 1.35 & 21.41 $\pm$ 0.08 \\ 
2014-03-13 & 56729.03 & 1080.49 & Spitzer & 3.6  & 480 & -  & 21.77 $\pm$ 0.27 \\
2014-03-13 & 56729.03 & 1080.49 & Spitzer & 4.5  & 480 & -  & 21.88 $\pm$ 0.24\\
\enddata
\tablecomments{Log of late time observations of {\em Swift} J1644+57 obtained with the {\em Hubble Space Telescope} and the {\em Spitzer Space Telescope}
Photometry is listed with and without host subtraction.}
\label{hst_spitzer}
\end{deluxetable}

\begin{deluxetable}{lllllll} 
\tablecolumns{8} 
\tablewidth{0pc}
\tablecaption{Late time X-ray observations of {\em Swift} J1644+57} 
\tablehead{ 
\colhead{Date-obs} & \colhead{MJD-obs} & \colhead{$\Delta T$} & \colhead{Telescope} & \colhead{ks} & \colhead{Count rate } & \colhead{Flux}  \\ 
& & \colhead{(days)} & &  & \colhead{(0.3-10 keV)} & \colhead{(erg s$^{-1}$ cm$^{-2}$)}  \\ }
\startdata
2012-09-27 &56197.81 &549.27 & XMM & 22.7& (1.9 $\pm$ 0.3) $\times 10^{-3}$  & 9.93 $\times 10^{-15}$ \\ 
2012-10-05  & 56205.80 &557.26 & XMM & 28.7 &  (1.2 $\pm$ 0.2) $\times 10^{-3}$ &6.27 $\times 10^{-15}$ \\
2012-11-26 & 56257.44 & 608.90 & Chandra & 24.7 & (3.0 $\pm$ 1.1) $\times 10^{-4}$ & $4.18 \times 10^{-15}$ \\ 
2013-07-17 & 56490.70 & 842.16 & XMM & 44.1 &  (8.1 $\pm$ 1.5) $\times 10^{-4}$ & $4.21 \times 10^{-15}$ \\
2015-02-17 & 57070.20 &1421.66 & Chandra & 27.8  & (4.6 $\pm$ 1.3) $\times 10^{-4}$ & $6.40 \times 10^{-15}$ \\
2015-04-06 & 57118.85 & 1470.31 & Chandra  & 18.7 & (2.7 $\pm$ 1.3) $\times 10^{-4}$ & $3.76 \times 10^{-15}$ \\
- & - & 524.32 & XRT & - & $(1.90 \pm 0.69) \times 10^{-4}$ & $9.18 \times 10^{-15}$  \\ 
- & - & 602.57 & XRT & - & $(1.61 \pm 0.67) \times 10^{-4}$ & $7.77 \times 10^{-15}$ \\
\enddata
\tablecomments{Log of late time observations of {\em Swift} J1644+57 obtained with {\em XMM-Newton}, {\em Chandra} and the {\em Swift} XRT.}
\label{xray}
\end{deluxetable}

\begin{deluxetable}{lll} 
\tablecolumns{8} 
\tablewidth{0pc}
\tablecaption{Host galaxy photometry for the host of {\em Swift} J1644+57} 
\tablehead{ 
\colhead{Band} & \colhead{Mag (AB)} & \colhead{Ref}}
\startdata
B & 24.14 $\pm$ 0.05 & \cite{levan11} \\
g &   23.66 $\pm$ 0.05 & \cite{levan11}\\
r &    22.80 $\pm$ 0.10 & \cite{levan11} \\
F606W   & 22.72 $\pm$ 0.03 & This work\\
i &  22.31 $\pm$ 0.10 & \cite{levan11}\\
z &   22.03 $\pm$ 0.03 & This work \\
J &   21.87 $\pm$ 0.06 & This work \\
H &  21.63 $\pm$ 0.04 & This work \\
F160W  & 21.53 $\pm$ 0.01  & This work \\
K &   21.42 $\pm$ 0.04 & This work\\
Spitzer ch1 &   21.77 $\pm$ 0.27 & This work \\
Spitzer ch2 &  21.88 $\pm$ 0.24 & This work \\
WISE W3 &  $>17.95$ & \cite{levan11}\\
WISE W4 &  $>16.14$  & \cite{levan11} \\
\enddata
\tablecomments{Since {\em HST} observations indicate
at early times there was a small transient contribution even in the  optical bands we have included an additional error of 0.1 mag on the $r$ and
$i-$band data. }
\label{host_tab}
\end{deluxetable}

\begin{figure}
    \centering
    \includegraphics[width=12cm,angle=90]{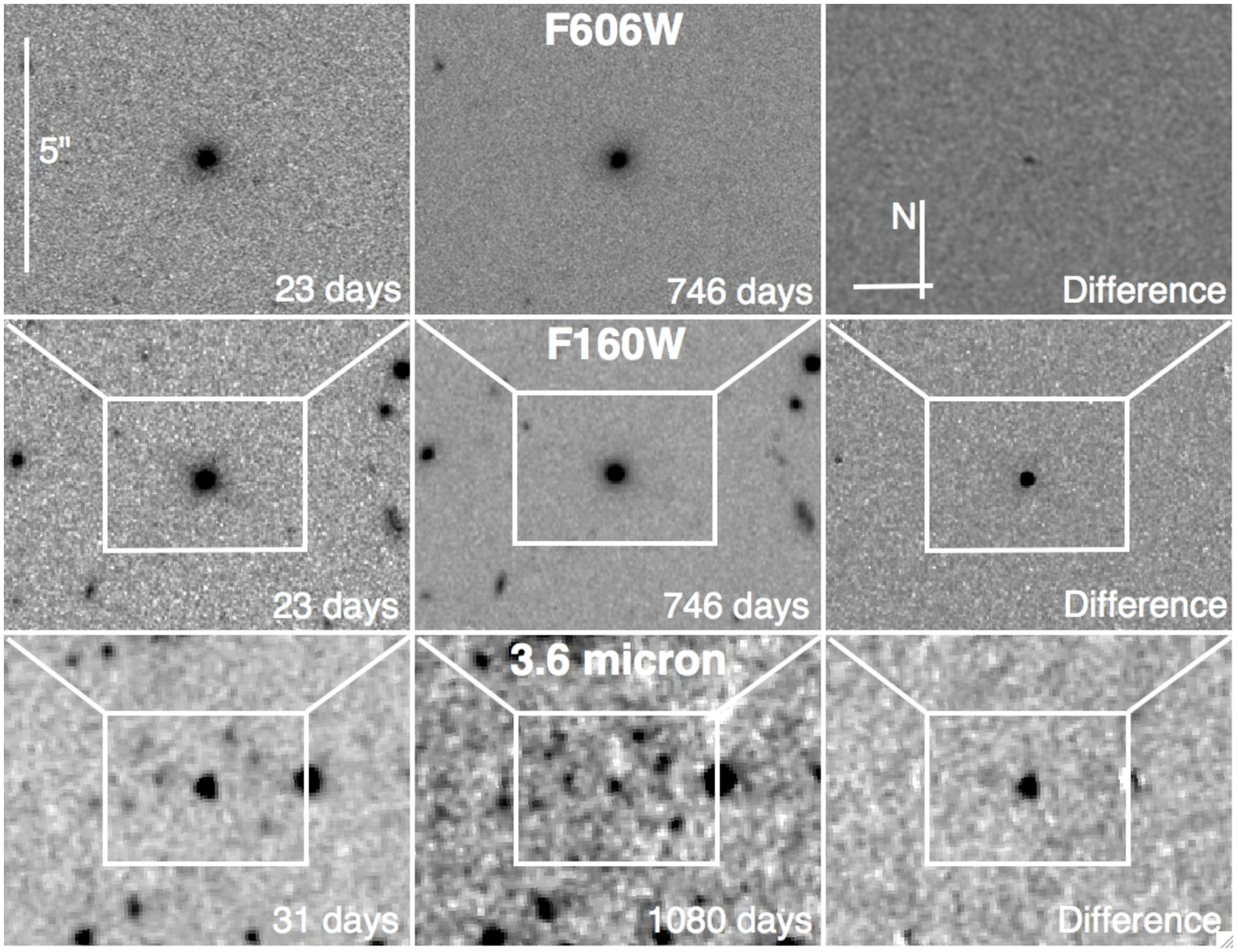}
\caption{Early to late time space based observations of {\em Swift} J1644+57 with {\em HST} and {\em Spitzer}. 
At early times the nIR and mid-IR are dominated by afterglow emission, while in the optical the host dominates at all epochs, although a weak transient
can be seen in our F606W observations.}
\label{mosaic}
\end{figure}

\begin{figure}
    \centering
    \includegraphics[width=8cm,angle=0]{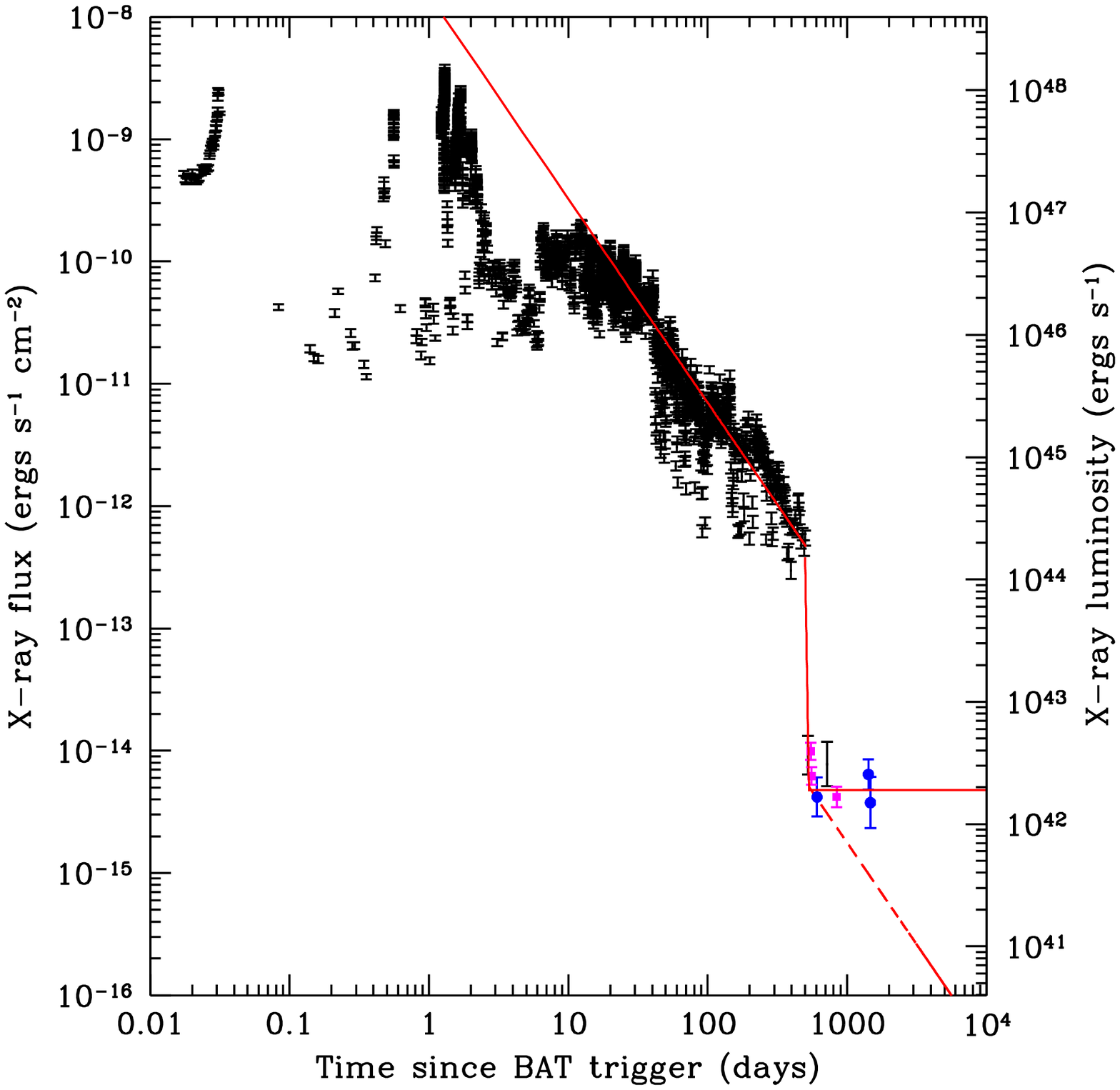}
     \includegraphics[width=8cm,angle=0]{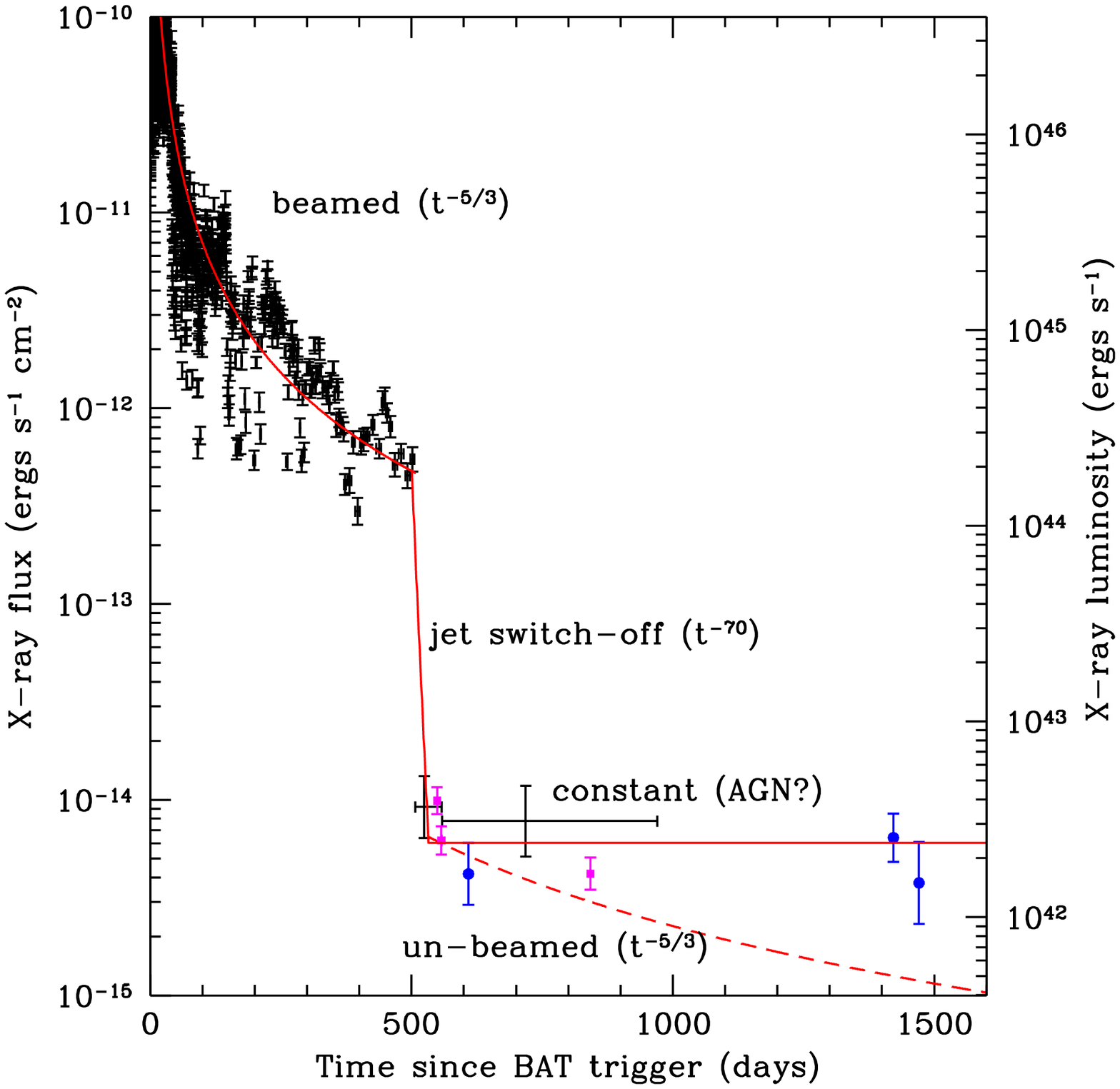}
\caption{The X-ray light curve of {\em Swift} J1644+57 obtained with the {\em Swift}-XRT (black), {\em XMM-Newton} (magenta) and {\em Chandra} (blue). The data plotted
in each fit are identical but are plotted on a logarithmic (left) and linear (right) scale to emphasise both the overall shape, and behaviour after the rapid decay. The solid red
line shows a $t^{-5/3}$ decay plotted through the X-ray observations. This is not a fit to the data, but an indicative reference model. A sharp break of $t^{-70}$ is shown
at 500 days, followed by a constant level. For comparison, a continued decay of $t^{-5/3}$ after the end of the steep decline is shown as the dashed line. }
\label{xlc}
\end{figure}

\begin{figure}
    \centering
    \includegraphics[width=8cm,angle=270]{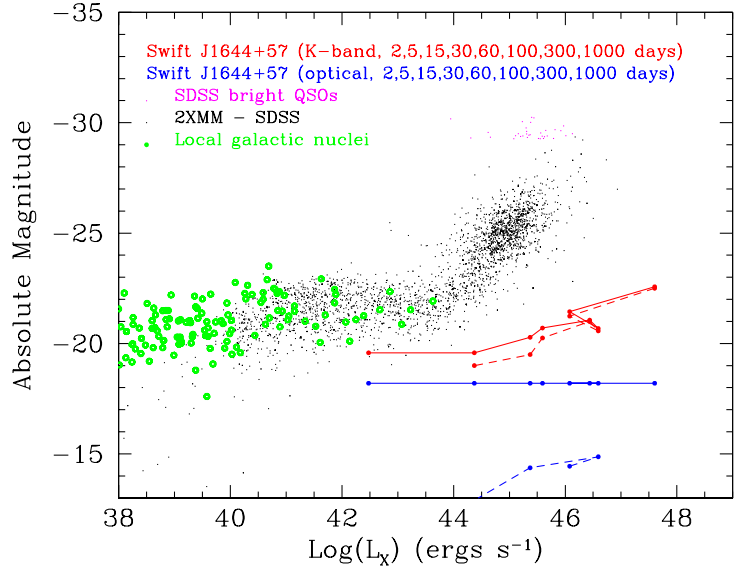}
\caption{The evolution of the location of the transient in the $L_{\rm X}$ -- $M_{\rm opt/IR}$ plane, showing the infrared and optical fading over several 
years following the first outburst. While at early times the source occupied a region of parameter space largely distinct from that of other transients, its
final location is much closer to the local of normal galaxies. However, it remains unusually X-ray luminous given its optical absolute magnitude. The solid lines represent
the total observed light (host galaxy plus transient), while the dashed lines show the host subtracted transient light, not-corrected for host galaxy extinction. }
\label{lxmopt}
\end{figure}

\begin{figure*}
    \centering
    \includegraphics[width=10cm,angle=270]{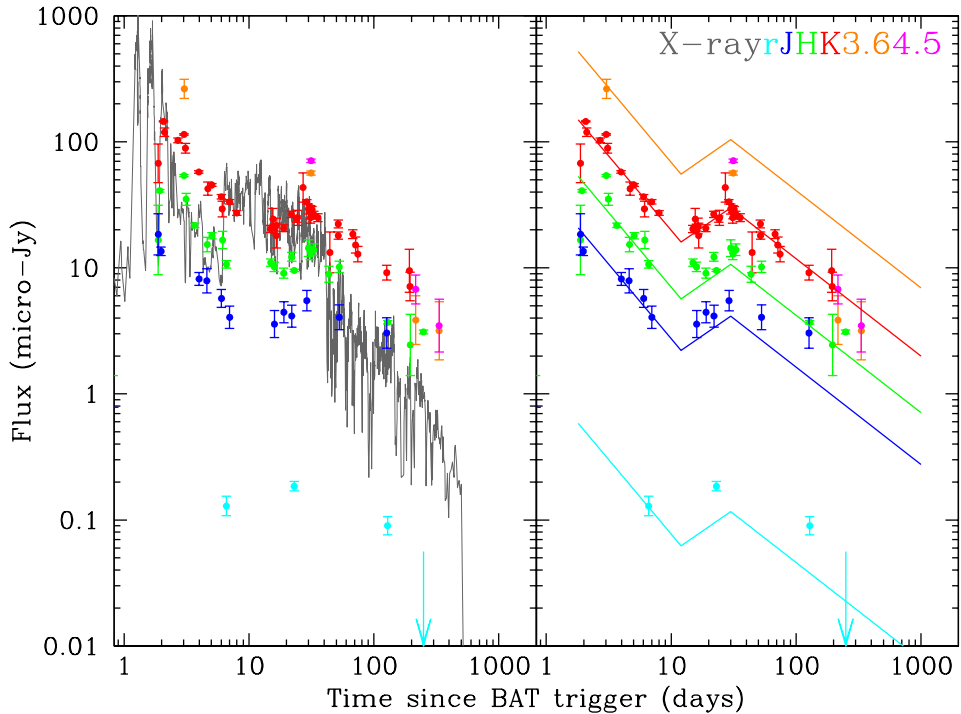}
\caption{
Optical and infrared photometry of {\em Swift} J1644+57 with the contribution of the host galaxy subtracted. 
The early time behaviour is apparently achromatic, with a constant offset between the bands up to $\sim 10$ days
after the BAT trigger, although some variability is visible on top of a gradual decay. After 10 days the counterpart
re-brightens to a bump that peaks 30-50 days after trigger. The left hand panel shows the X-ray light curve for comparison, although
there is a significant re-brightening in X-ray's it occurs well before the optical/IR brightening, and is much sharper. 
The optical bump feature is also shown in all the available bands, although is clearly stronger in the bluer bands. 
After the peak the behaviour is apparently chromatic, with the redder bands falling more rapidly than the blue. }
\label{optlc}
\end{figure*}

\begin{figure}
    \centering
    \includegraphics[width=12cm,angle=0]{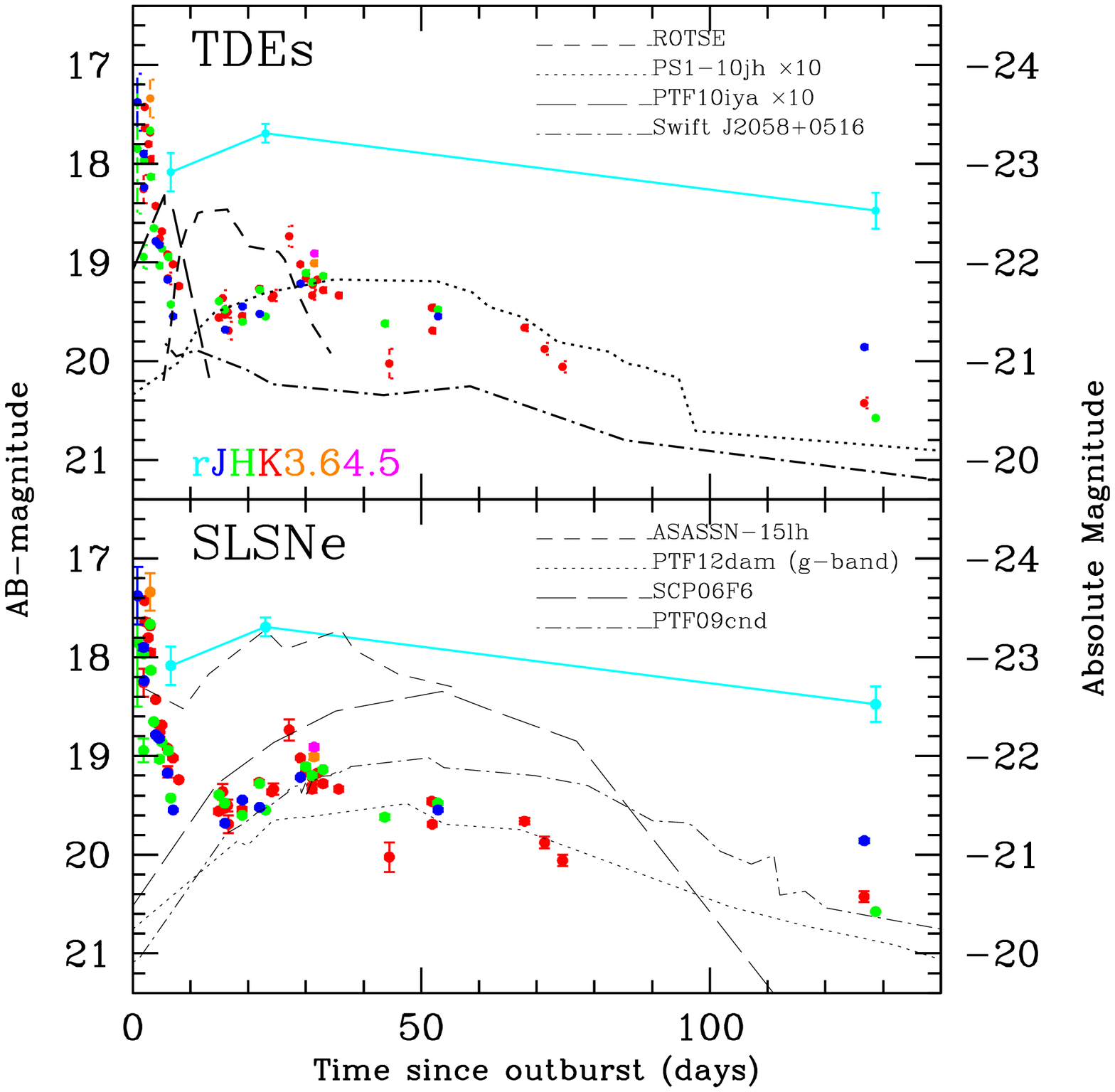}
\caption{A comparison of the host subtracted, extinction corrected light curves of {\em Swift} J1644+57 with other luminous transient events, 
in particular the light curves of suggested candidate tidal disruption systems (top, including the luminous ``Dougie" discovered by ROTSE \citep{vinko14}, PS1-10jh \citep{gezari12}, PTF10iya \citep{cenko11} and {\em Swift} J2058+0516 \citep{cenko12,pasham15}) and SLSNe (bottom, including ASASSN-15lh \citep{dong15}, PTF12dam \citep{nicholl13}, SCP06F6 \citep{barbary09} and PTF09cnd \citep{quimby11}). Unfortunately, the poor sampling of the
optical component of {\em Swift} J1644+57 (cyan line) makes a direct comparison with the predominantly optical observations of other transient
classes difficult. However, SLSNe can provide a reasonable match to the observations (in particular ASASSN-15lh), while TDFs match the light curve
shape, but are required to be significantly brighter than previous examples. }
\label{sncomp}
\end{figure}

\begin{figure}
    \centering
    \includegraphics[width=8cm,angle=0]{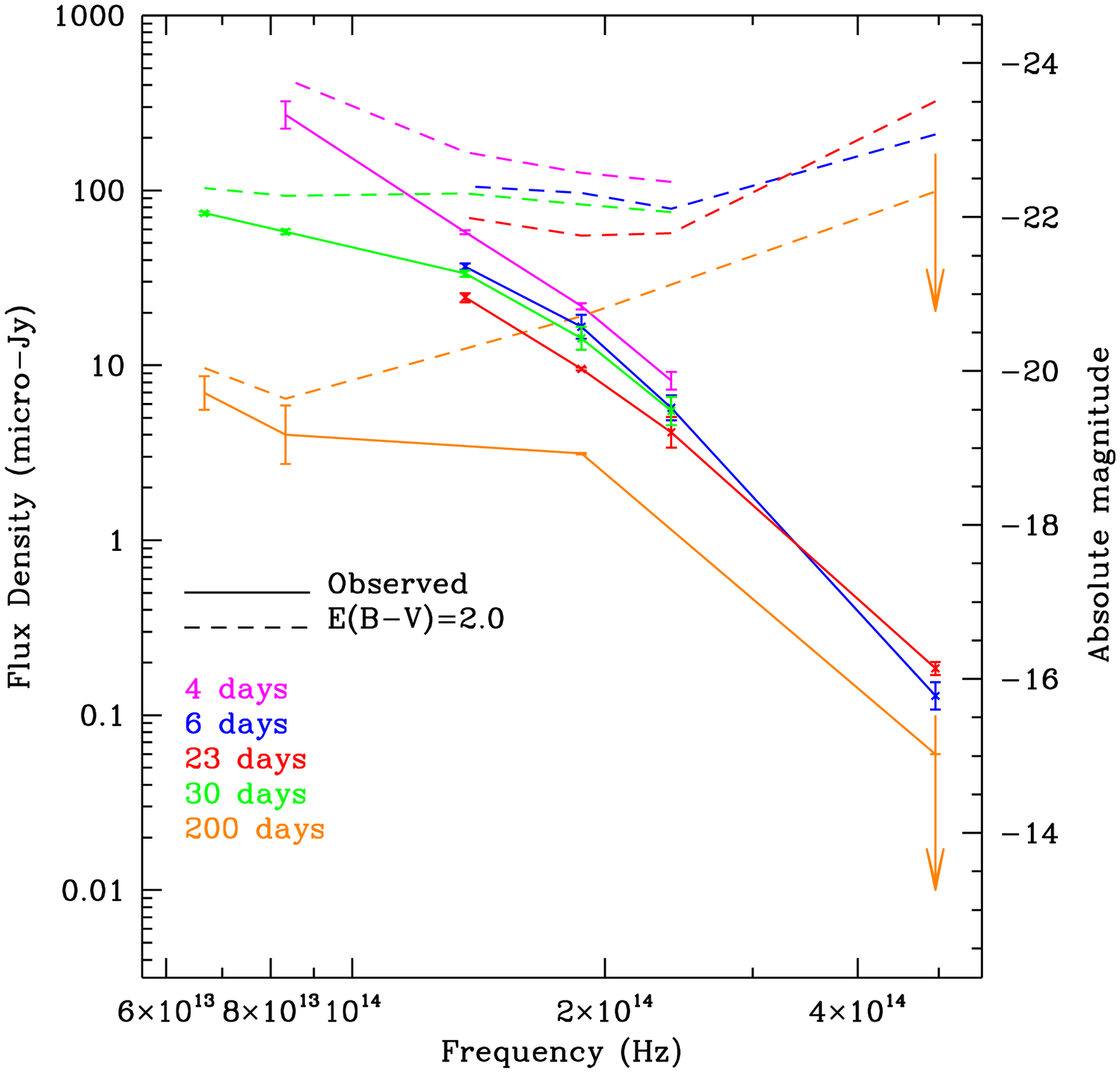}
     \includegraphics[width=8cm,angle=0]{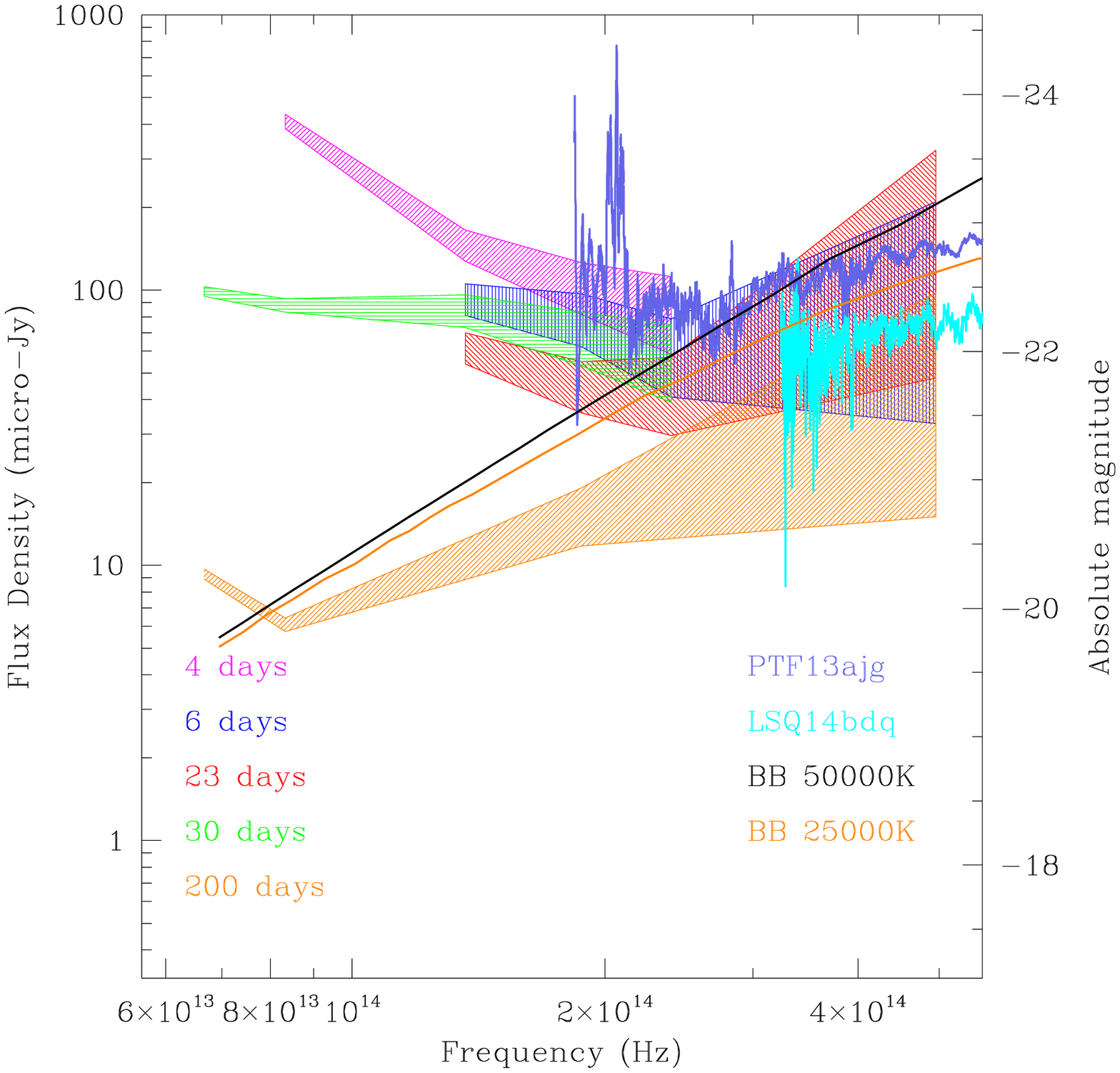}
\caption{The spectral energy distribution of the {\em Swift} J1644+57 at 4 representative epochs spanning the 200 days after outburst. The left hand panel shows the
multiple epochs as observed (solid lines) and corrected for $E(B-V)_{\mathrm{host}} = 2$ (dashed lines).  The right hand panel shows the extinction correct SED (the
shaded region represents the range between $E(B-V)_{\mathrm{host}} = 2$ and $E(B-V)_{\mathrm{host}} = 1.5$) in comparison with two representative lines of SLSNe 
in particular PTF13ajg (blue) and PTF14bdq (cyan), taken from WISEREP \citep{WISEREP}, as
well as black bodies of two different temperatures.}
\label{optsed}
\end{figure}

\begin{figure}
    \centering
    \includegraphics[width=12cm,angle=90]{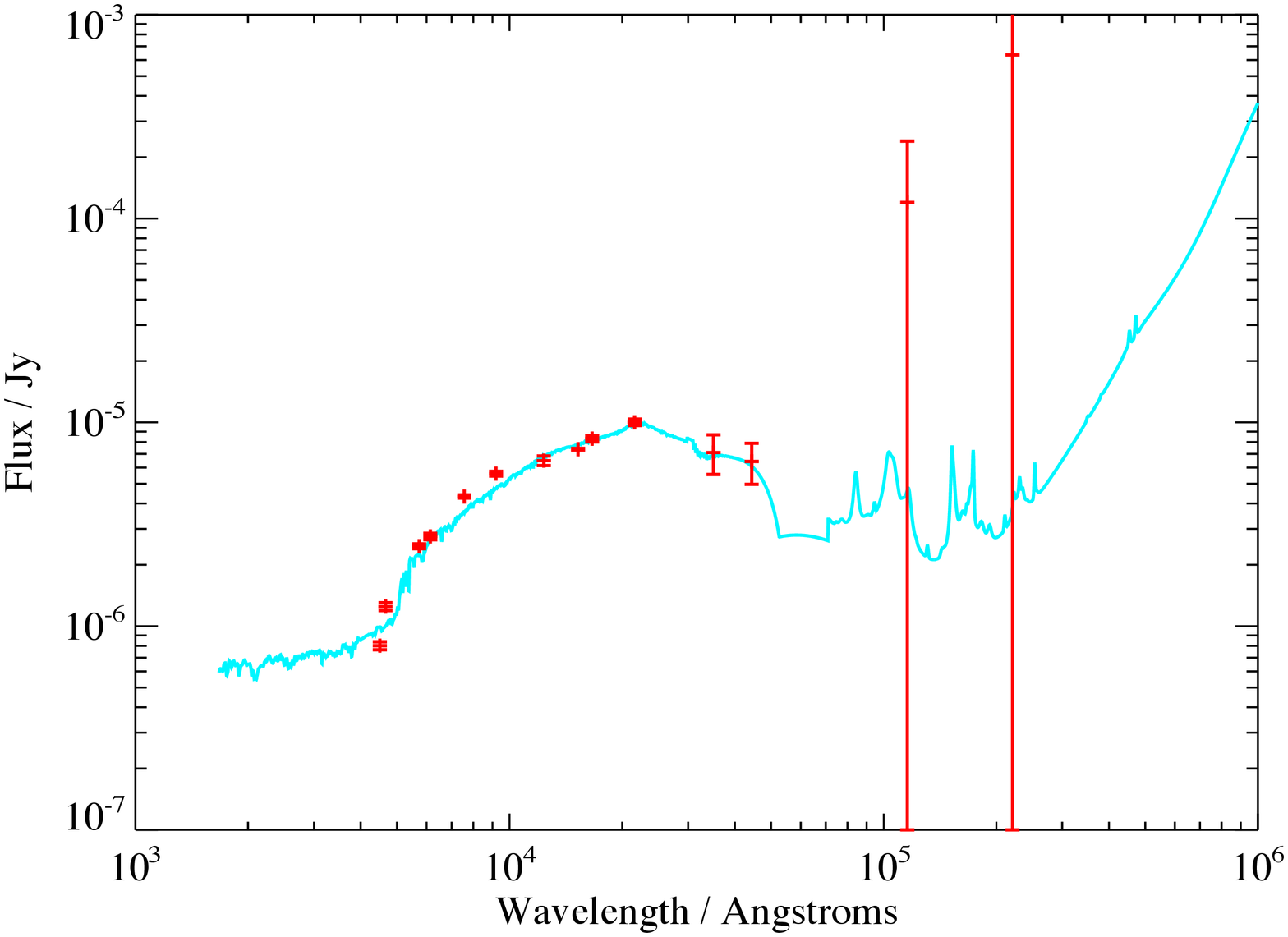}
\caption{The spectral energy distribution of the host galaxy of {\em Swift} J1644+57 from our late time
photometry, together with our best-fitting BPASS model. 
The relatively red optical colours favour a  system dominated by an older underlying population, consistent with a morphological classification as an elliptical galaxy. 
However, emission lines observed in the optical spectrum demonstrate the presence of  some ongoing star formation.  
}
\label{galsed}
\end{figure}

\end{document}